\documentclass{aa}  

\usepackage{graphicx}
\usepackage{txfonts}
\usepackage{natbib}

\usepackage[colorlinks=true, linkcolor=blue, citecolor=blue]{hyperref}
\usepackage{longtable}
\usepackage[switch]{lineno}

\begin{document}

   \title{Evaluating the chromospheric structure model of AD Leo using RH 1.5D and magnetic field data}

    \author{Shuai Liu\inst{1,2}
       \and Jianrong Shi\inst{1,3,4}\thanks{Corresponding author: \email{sjr@nao.cas.cn}}
       \and Huigang Wei\inst{1}\thanks{Corresponding author: \email{whg@nao.cas.cn}}
       \and Wenxian Li\inst{1,5}
       \and Jifeng Liu\inst{1,3}
       \and Shangbin Yang\inst{3,1}
       \and Henggeng Han\inst{1}
    }
    
    \institute{
    Key Laboratory of Optical Astronomy, National Astronomical Observatories, Chinese Academy of Sciences, Beijing 100011, PR China    
    \and
    Instituto de Astrofísica de Canarias, Vía Láctea, 38205 La Laguna, Tenerife, Spain
    \and
    School of Astronomy and Space Science, University of Chinese Academy of Sciences, Beijing 100049, PR China
    \and
    School of Physics and Technology, Nantong University, Nantong 226019, China
    \and
    Key Laboratory of Solar Activity and Space Weather, National Space Science Center, Chinese Academy of Sciences, Beijing 100190, PR China
    }

  \abstract
    % Context
    {The interplay between surface magnetic topology and chromospheric heating in active M dwarfs remains poorly constrained, limiting our understanding of their magnetic cycles and high‐energy environments.}
    % Aims
    {We aim to test whether detailed zeeman–doppler imaging (ZDI) maps of AD Leo can be used to spatially anchor a multi‐component chromospheric model and thus validate the link between magnetic flux distribution and emission‐line formation.}
    % Methods
    {We analyze high-resolution CARMENES spectra of H$_\alpha$ and the \ion{Ca}{II} infrared triplet, together with detailed ZDI maps. The RH1.5D non-local thermodynamic equilibrium (non-LTE) radiative transfer code is employed to synthesize spectral lines with two active atmospheric components (low-latitude near the equator and polar near the pole) and a quiet background. Their relative filling factors and temperature structures are optimized per epoch. The ZDI maps serve as a qualitative reference for the large-scale magnetic topology but are not used as input to the optimization procedure.}
    % Results
    {Our model reproduces the spectral line profiles over multiple epochs. The low-latitude active region shows notable variability—accounting for approximately 55–86\% of the emission, while the polar active region remains relatively constant in area (12–17\%) but exhibits temperature variations over time, particularly during the periods of increased activity. The spatial locations of the active regions derived from spectroscopy are in good agreement with the radial magnetic field distribution obtained from ZDI.}
    % Conclusions
    {Our results indicate that combining spectroscopic modeling with magnetic field maps is an effective approach for mapping magneto‑chromospheric structures in M dwarfs. This framework deepens our understanding of stellar magnetic cycles and chromospheric dynamics, paving the way for detailed time‑resolved studies in active low‑mass stars.}


   \maketitle

\section{Introduction} \label{sec:intro}

Magnetic activity in M dwarfs plays a vital role in shaping their dynamic chromospheric behavior and can significantly influence the radiation environment of orbiting planets. Gaining insight into the coupling between magnetic fields and upper atmospheric layers is key to understanding their stellar atmospheres and potential impact on exoplanet habitability\citep{2013A&A...557A..67V,2024ARA&A..62..593H}.
Among the spectral features shaped by magnetic fields in M dwarfs, lines such as H$_\alpha$ and the \ion{Ca}{II} infrared triplet (IRT) serve as prominent diagnostics of chromospheric behavior.
Their profiles are intricately affected by  non-local thermodynamic equilibrium (non-LTE) radiative processes and magnetic topology \citep{1996AJ....112.2799H,2017ApJ...834...85N}.
Previous studies have modeling chromospheric properties using spectral analysis, but these approaches are limited by the absence of high-resolution magnetic field maps \citep{1998A&A...336..613S,2005A&A...439.1137F,2019A&A...623A.136H}.
Early models by \citet{1982ApJ...258..740G} demonstrated that a single-component chromospheric model could independently fit the \ion{Ca}{II} and H$_\alpha$ lines; however, such models rely on assumptions of extremely low microturbulent velocities (less than \( 2\ \mathrm{km\ s}^{-1} \)). These assumptions are inconsistent with observations of solar and stellar chromospheres, where microturbulent velocities typically increase with height, reaching values of \( 5-10\ \mathrm{km\ s}^{-1} \) in the upper chromosphere \citep{2005A&A...439.1137F}. Furthermore, as noted by \citet{2020PhDT.........7H}, single-component models fail to simultaneously reproduce other chromospheric diagnostics, such as the \ion{Mg}{II} \textit{h}\&\textit{k} lines, particularly at higher activity levels. Modern solar and stellar models suggest that the chromosphere is highly structured and dynamic, with significant spatial and temporal inhomogeneities \citep{1992ApJ...397L..59C,1995ApJ...440L..29C}. These effects strongly influence line formation, making a single-component model inadequate for reproducing the observed time-resolved spectral features.
The advent of sophisticated observational techniques \citep{2008MNRAS.390..567M, 2023A&A...676A..56B} has enhanced our ability to explore the relationship between magnetic fields and chromospheric activity, leading to improved predictions of stellar variability and planetary habitability.

In our previous study \citep{2024ApJ...975..133L}, we used the high-resolution spectra to model chromospheric activity in the M dwarf G\,80-21. We found two distinct regions: —a small polar active region and a large surface-spanning one, however we could not confirm their spatial distribution due to the lack of magnetic field data.

AD\,Leo, the target of this study, provides an opportunity to overcome this limitation. It is a well studied M dwarf with strong chromospheric emission and well documented magnetic field observations. AD Leo is a highly active M dwarf star with a rapid rotation period of 2.23 days \citep{2008MNRAS.390..567M,2012PASP..124..545H,2023A&A...674A.110C}. It has also shown long-term variability and \citet{2014ApJ...781L...9B} reported a roughly 7-year activity cycle based on the ASAS photometry and CASLEO spectroscopy data. Several studies have modeled AD\,Leo's chromosphere. \citet{1994A&A...281..129M} derived a semi-empirical model of the quiescent state by fitting continuum and chromospheric lines. \citet{1992ApJS...78..565H} constructed a grid of flare models, including a quiescent model, based on the photosphere and incorporating X-ray heating from the corona.  \citet{1998A&A...336..613S} used a linear temperature-log mass relation to model the chromosphere and transition region.  \citet{2005A&A...439.1137F} modeled AD\,Leo's chromosphere using PHOENIX model, focusing on the H$_\alpha$, Na\,D, and various Fe and Mg lines in the region of 3700–3900\,\AA;
however, they only provided an approximate fit to the observed spectra. The work of \citet{2008MNRAS.390..567M} mapped its magnetic field with a stable, dipolar geometry. In contrast, the recent study by \citet{2023A&A...676A..56B} revealed significant changes in its magnetic field over 14 years, with a reduction in axial symmetry from 99\% to 60\%, indicating a more complex magnetic geometry. These magnetic field maps allow us to directly test the correspondence between magnetic structures and the chromospheric features we inferred from spectral data. The inclusion of magnetic field maps in this study allows us to test whether the magnetic field, particularly the radial component, contributes to the inferred chromospheric structure.

In this study, we aim to refine our understanding of the chromospheric activity in AD\,Leo by combining the high-resolution spectra with the magnetic field maps. Specifically, we test whether the small polar and large surface active regions inferred from our spectral analysis correspond to the areas of concentrated magnetic field. This will provide a critical test of our previous model by incorporating magnetic field constraints, which are absent from the earlier study.
In Section~\ref{sec:obs}, we describe the high-resolution CARMENES spectra of AD\,Leo, focusing on the H$_\alpha$ and \ion{Ca}{II} IRT lines. In Section~\ref{sec:mod}, we explain our chromospheric model construction using extended MARCS-OS atmospheres and non-LTE computations with MULTI and RH1.5D. In Section~\ref{sec:result}, we analyze the effects of magnetic fields on the synthetic spectra and their correlation with active regions. Finally, Section~\ref{sec:con} summarizes our conclusions.

\section{Observations} \label{sec:obs}
\subsection{Observations and data reduction} 

We analyzed 46 high–resolution spectra of AD Leo taken with the VIS channel of the CARMENES spectrograph on the 3.5 m Calar Alto telescope \citep{2018SPIE10702E..0WQ}. The VIS arm covers 5200–9600 $\AA$ at \(R\approx94\,600\), providing signal‐to‐noise ratios of 13–69 in the H\(_\alpha\) region and 25–111 in the \ion{Ca}{II} IRT windows. Observations span from 2018 March 21 to 2020 February 16, with airmass values between 1.05 and 2.33. Fig.~\ref{figure1} (top panels) displays these time‐resolved spectra of AD Leo, where the shaded yellow band indicates the fitting interval and colored traces correspond to sequential epochs. Four spectra with exceptionally strong emission are highlighted in black. We supplemented these data with contemporaneous zeeman–doppler imaging (ZDI) maps from \citet{2023A&A...676A..56B} to relate chromospheric emission to surface magnetic‐flux distributions.

The ZDI maps adopted in this study were reconstructed by \citet{2023A&A...676A..56B} using spectropolarimetric observations of AD\,Leo obtained with ESPaDOnS and Narval in the optical (2006--2019), and with SPIRou in the near-infrared (2019--2020).
For our analysis, we focus on the SPIRou observations collected between February 2019 and June 2020, subdivided into four epochs (2019a–2020b) to ensure magnetic field stability within each interval. These epochs each span 15–30 stellar rotations and provide good phase coverage. The ZDI maps were reconstructed using least-squares deconvolution (LSD) profiles derived from 3240 spectral lines in the optical and 838 lines in the near-infrared, after excluding regions affected by strong telluric absorption. The observational properties of the SPIRou-based epochs, including date ranges, number of observations, phase coverage, and average S/N, are summarized in Table~\ref{table:combined}

\begin{table*}[ht]
\centering
\caption{Summary of CARMENES and ZDI Observations (from \citet{2023A&A...676A..56B}) for AD Leo}
\label{table:combined}
\begin{tabular}{lccccc}
\hline
\multicolumn{6}{c}{CARMENES Observations} \\
\hline
Epoch & Date Range & Observations & Phase Coverage (cycles) & S/N (\ion{Ca}{II}) & S/N ($H_\alpha$) \\
\hline
2018 & 2018-03-21 to 2018-04-17 & 26 & 12 & 13.6–69.38 & 25.84–111.05 \\
2020 & 2020-02-13 to 2020-02-16 & 20 & 1 & 26.54–55.01 & 45.58–85.88 \\
\hline 
\multicolumn{6}{c}{ZDI Observations} \\
\hline
Epoch & Date Range & Observations  & Phase Coverage (cycles) & \multicolumn{2}{c}{S/N (ZDI)} \\
\hline
2019a & 2019-04-15 to 2019-06-21 & 21 & 30 & \multicolumn{2}{c}{130–193} \\
2019b & 2019-10-16 to 2019-12-12 & 21 & 26 & \multicolumn{2}{c}{68–201} \\
2020a & 2020-01-26 to 2020-03-12 & 30 & 20 & \multicolumn{2}{c}{166–218} \\
2020b & 2020-05-08 to 2020-06-10 & 18 & 15 & \multicolumn{2}{c}{120–217} \\
\hline
\end{tabular}
\end{table*}

Each spectrum was continuum‐normalized following \citet{2019A&A...623A..24F}, using two narrow bands bracketing each target line. For H\(_\alpha\), the reference intervals were 6547.4–6557.9 $\AA$ and 6577.9–6586.4 $\AA$; for the \ion{Ca}{II} IRT lines, 8476.3–8486.3 $\AA$ and 8552.4–8554.4 $\AA$. The median flux in each band was averaged to define a local continuum level, and the observed flux was divided by this value to place all epochs on a uniform relative scale.

\subsection{Stellar parameters and quiet‐region template} \label{sec:params}

\begin{figure*}[htbp]
\includegraphics[width=0.95\hsize]{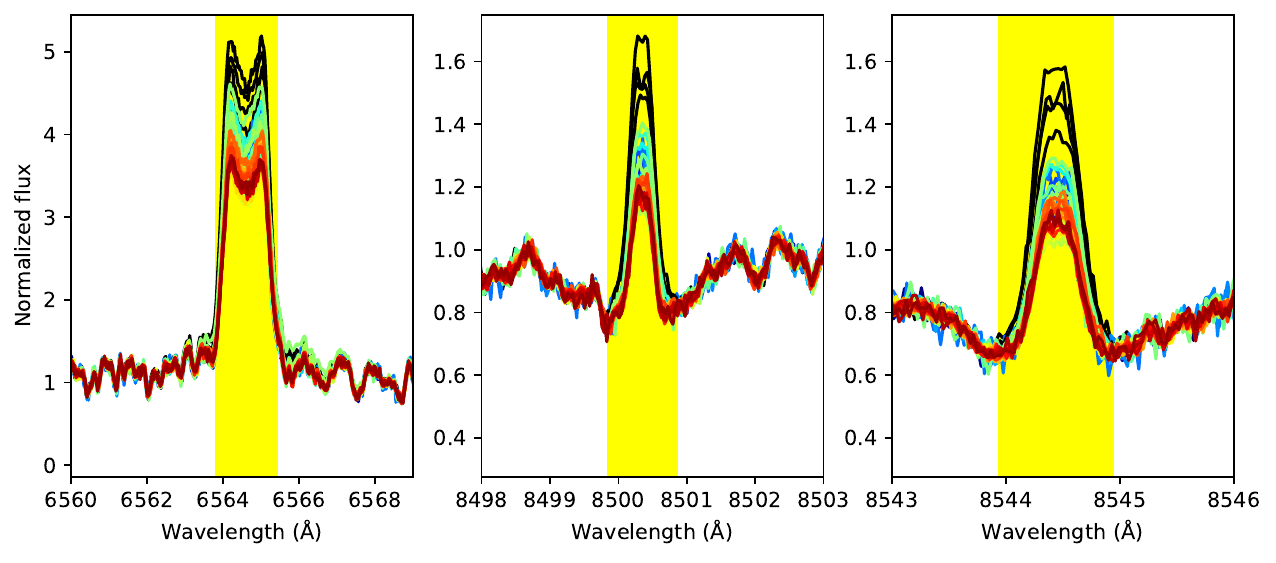}
\includegraphics[width=0.95\hsize]{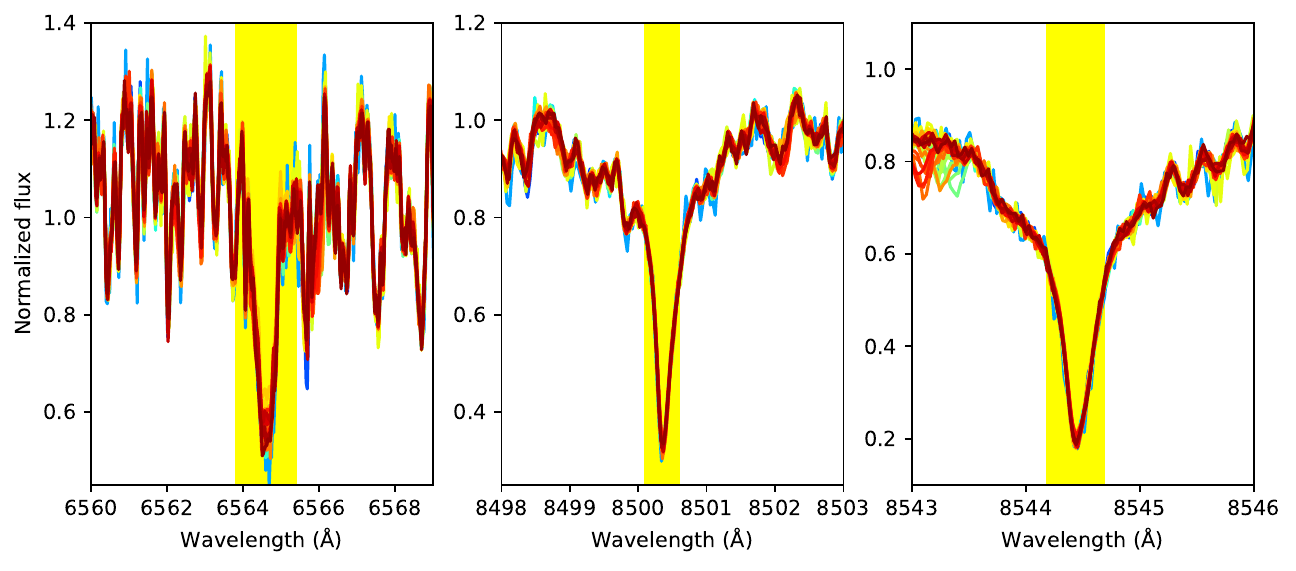}
\includegraphics[width=0.95\hsize]{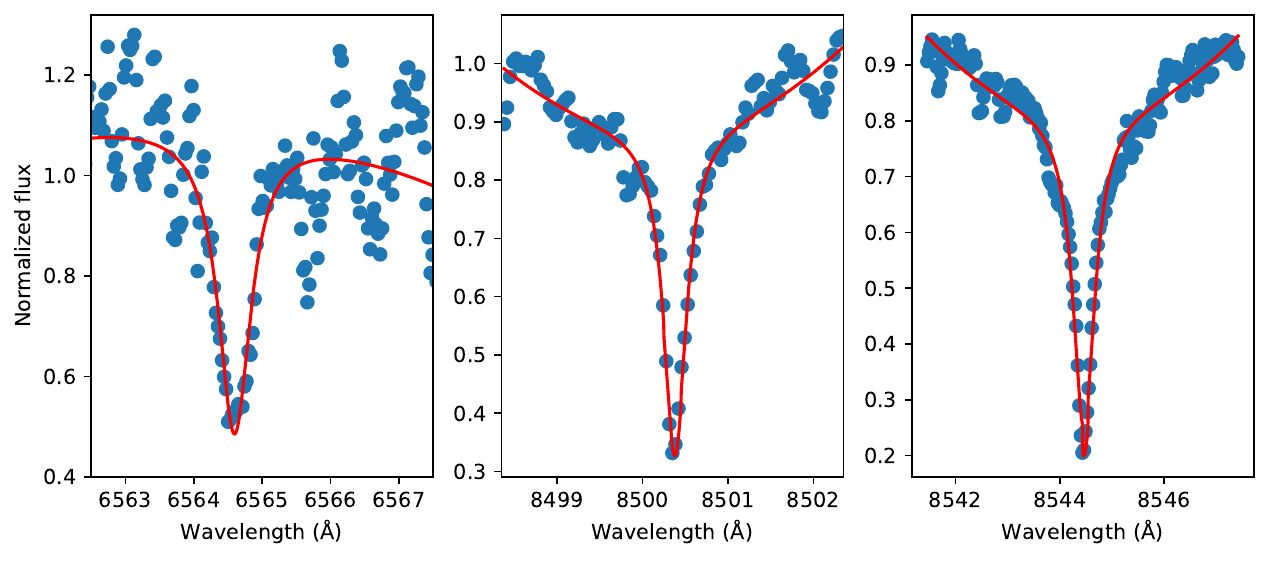}
\caption{Observations of AD Leo (top panels) and of the inactive template LP 776-46 (middle panels). All spectra have been continuum‐normalized consistently across epochs. The shaded yellow region marks the wavelength interval used for model fitting. Colored traces correspond to sequential observations, progressing from red to blue, with four exceptionally strong emission events in AD Leo highlighted in black. The bottom panels show the highest-SNR time-resolved spectra (SNR$\sim$98 for H\,$\alpha$ and 150 for \ion{Ca}{II} IRT) of the inactive template star LP 776-46, with overlaid Lorentzian profile fits (red curves).}

\label{figure1}
\end{figure*}

We adopt \(T_{\rm eff}=3477\pm23\)\,K, \(\log g=5.12\pm0.12\), and \([\rm Fe/H]=-0.19\pm0.12\) for AD Leo from \citet{2021A&A...656A.162M}.
As an inactive reference, we constructed a quiet‐region template from LP 776-46 (\(T_{\rm eff}=3464\pm27\)\,K, \(\log g=5.08\pm0.10\), \([\rm Fe/H]=-0.13\pm0.09\)), using its time‐series spectra (Fig.~\ref{figure1}, middle panels) co‐added to achieve SNR $\sim$ 98 in H\(_\alpha\) and 150 in the \ion{Ca}{II} IRT. These high‐SNR observations were fitted with Lorentzian profiles in both H\(_\alpha\) and the \ion{Ca}{II} IRT bands to define the baseline for our active-region modeling (Fig. \ref{figure1}, bottom panels).

\section{Chromospheric model construction}\label{sec:mod}

Our modeling builds on the hybrid 1.5D framework introduced in \citet{2024ApJ...975..133L}, tailored here to AD Leo’s atmospheric parameters. We begin with a MARCS-OS photospheric structure \citep{2008A&A...486..951G} matching \(T_{\rm eff}\), \(\log g\), and [Fe/H] from \citet{2021A&A...656A.162M}, then insert three schematic temperature rises representing the lower chromosphere, upper chromosphere, and transition region. These inserts are specified at certain column-mass depths with linear temperature gradients, expanding the original 57-layer grid to 105 layers. This refined vertical sampling ensures that steep thermal gradients are well resolved.

Non-LTE level populations and radiation fields are obtained in two stages. First, the MULTI code \citep{1985JCoPh..59...56S,1985ASIC..152..189S} computes statistical equilibrium for hydrogen and \ion{Ca}{II}, including partial redistribution effects. Departure coefficients derived from MULTI calculations are input into RH1.5D \citep{2015A&A...574A...3P}, which performs detailed line-profile synthesis including backwarming, opacity effects, and lateral averaging to approximate 3D radiative transfer effects. Throughout this process, we enforce consistency with AD Leo’s large-scale magnetic topology by adopting disc-integrated filling factors—geometric weights representing the fractional coverage of active regions—which are qualitatively consistent with the magnetic structures revealed by contemporaneous ZDI maps \citep{2023A&A...676A..56B}.

Rotational Doppler effects have been incorporated by applying a rotational convolution with the measured $v\sin i\approx3~\mathrm{km\,s^{-1}}$ \citep{2023A&A...674A.110C} to the synthetic spectra before comparison with the observations. Given this low projected rotational velocity, no additional differential Doppler shifts for individual active regions are expected to have a measurable impact on the integrated line profiles.

The chromospheric structure is described by six free parameters—two column‐mass depths ($m_{\min}, m_{\rm mid}$), two temperatures ($T_{\rm mid}, T_{\rm top}$), and the transition‐region slope ($\rm{grad}_{\rm TR}$). A uniform grid with even moderate sampling (e.g.,\ ten points per parameter) would require $10^6$ models, which is computationally expensive \citep{2013A&A...553A...6H}. Therefore, we adopt a two‐step grid‐search strategy. In the first step, we perform a coarse scan over broad, physically motivated parameter ranges (Table~\ref{table:set}) by computing synthetic H$\alpha$ and \ion{Ca}{II} IRT profiles for each grid point and evaluating a $\chi^2$ metric against the observed spectra, identifying the minimum $\chi^2$ as the initial best‐fit. In the second step, centered on this coarse‐scan minimum, we construct a finer sub‐grid in each parameter and repeat the $\chi^2$ evaluation to refine our estimates of $(m_{\min}, m_{\rm mid}, T_{\rm mid}, T_{\rm top}, \mathrm{grad}_{\rm TR})$.

In this work, the “filling factor” refers to the fractional surface coverage of each atmospheric component on the projected visible stellar disc. These filling factors are purely geometric weights—applied uniformly with height—from the low-latitude and polar active regions and the quiet background, summing to unity at each epoch.

\section{Results and discussions}\label{sec:result}

Our multi-component model successfully reproduces the observed H$\alpha$ and \ion{Ca}{II} IRT line profiles across all epochs. The derived filling factors for the two active components reveal a clear spatial and temporal structure, which we now examine in the context of the star's magnetic topology.

\begin{figure*}[htbp]
\includegraphics[width=1.0\hsize]{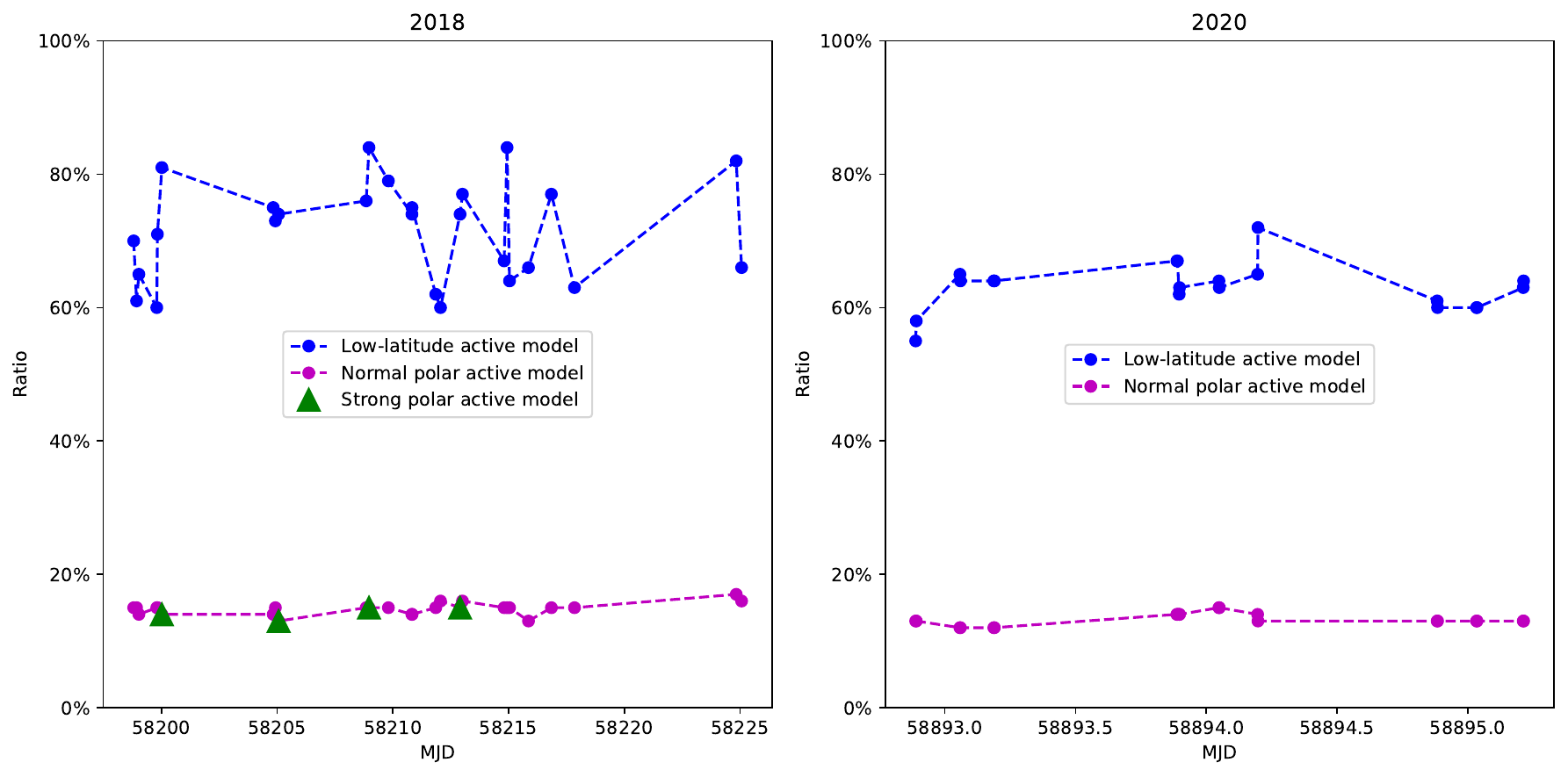}
\caption{Temporal evolution of filling factors during two observational campaigns. Left panel (2018): the low-latitude filling factor varies between 60\% and 86\%, while the polar component ranges from 12\% to 17\%. Right panel (2020): the low-latitude filling factor remains stable between 55\% and 72\%, with the polar component consistently contributing 12--17\%.}
\label{figure4}
\end{figure*}

\begin{figure*}
  \centering
  \includegraphics[width=0.9\textwidth]{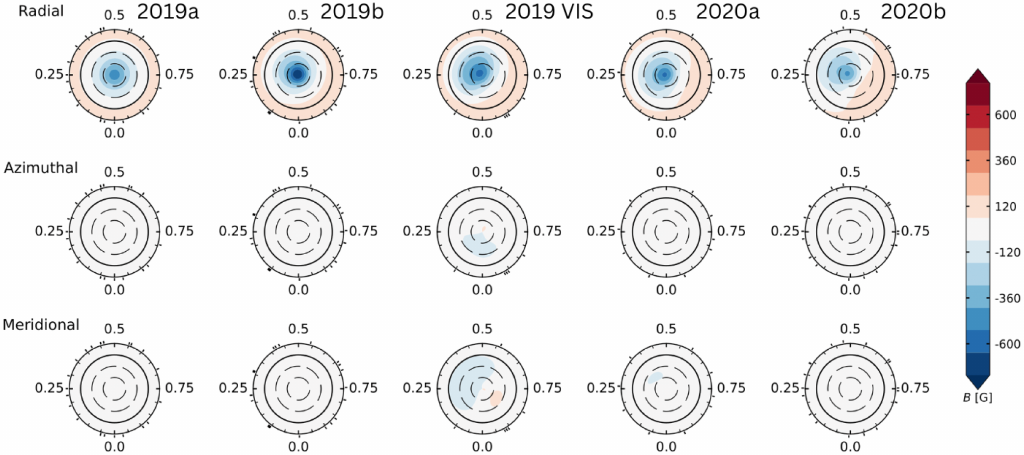}
  \caption{ZDI maps of AD Leo show the magnetic field components (radial, azimuthal, and meridional) in a flattened polar view across the epochs: 2019a, 2019b, 2019 optical, 2020a, and 2020b. Each component is color-coded to indicate magnetic polarity (red for positive, blue for negative). Radial ticks indicate the rotational phases during observations, and concentric circles represent stellar latitudes at $-30^\circ$, $+30^\circ$, $+60^\circ$, and the equator. The color bar is scaled to the maximum magnetic field strength for each epoch. For a stellar inclination of $i \approx 20^\circ$, latitudes south of $-20^\circ$ are not visible; the $-30^\circ$ latitude line is shown as a plotting reference only. Credit:\citet{2023A&A...676A..56B} } 
  \label{fig:zdi_maps}
\end{figure*}

Although we do not perform a latitude-resolving inversion, our interpretation aligns with the large-scale magnetic topology revealed by ZDI maps \citep[][see Fig. \ref{fig:zdi_maps}]{2023A&A...676A..56B}, which show more persistent radial fields near the poles and stronger variability at mid-to-low latitudes.
We note that the $-30^\circ$ latitude line shown in the flattened-polar ZDI maps is a plotting reference; for a stellar inclination of $i \approx 20^\circ$ \citep{2023A&A...674A.110C}, latitudes south of $-20^\circ$ are not visible and thus not directly constrained by the observations.
We acknowledge that spatial localization in our 1.5D framework is model-dependent and subject to projection ambiguities; precise latitudinal extents cannot be uniquely determined here.

\begin{figure}[htbp]
\includegraphics[width=1.0\hsize]{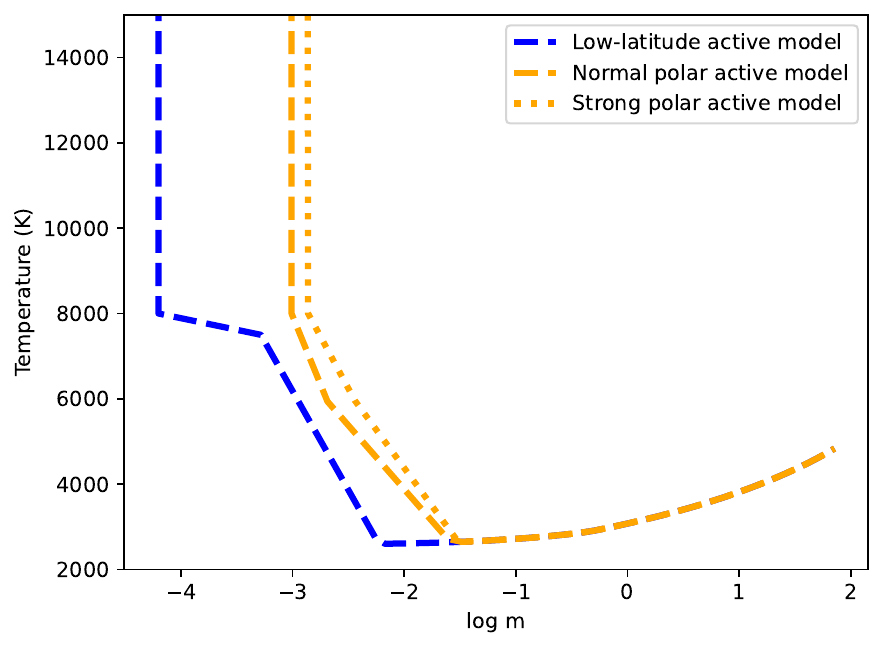}
\caption{Temperature stratification across chromospheric models: the plot illustrates how temperature varies with column mass (log m) for the low-latitude active model (dashed blue line) and for the polar models in both their normal (dashed orange) and enhanced (dotted orange) states.}

\label{figure2}
\end{figure}

This high level of stellar activity is manifested through frequent and energetic flares, which have been well documented in various studies \citep{2020A&A...637A..13M,2020PASJ...72...68N}. \citet{2020A&A...642A..53D} reported an inverse relationship between H$_\alpha$ and \ion{Ca}{II} emissions in AD Leo, suggesting a timescale-dependent correlation.

\subsection{Chromospheric temperature structures}\label{sec:structures}

Fig.~\ref{figure2} shows the temperature versus column-mass profiles for our two active regions. The low-latitude region (dashed blue) features a steep temperature gradient in the lower chromosphere, transitioning to a plateau in the upper chromosphere. This stratification closely resembles the solar VAL-C empirical model \citep{1981ApJS...45..635V}. In contrast, the polar region (solid orange) exhibits a more gradual temperature increase with a complex behavior before the transition region. 

These two distinct temperature structures are necessary to reproduce the full range of observed H\(\alpha\) core shapes (double-peaked versus single-peaked) and \ion{Ca}{II} infrared triplet intensities across our time-series spectra. This is consistent with our previous findings for G\,80$-$21 \citep{2024ApJ...975..133L}.

Importantly, our modeling procedure—applying a consistent stepwise fitting approach to both the low-latitude and polar components—revealed that only the polar component required an overheated structure to match the four spectra with the strongest emission. This enhanced polar heating was therefore a result derived from the fitting process, not an initial assumption.

All observed spectra can be modeled as a combination of the low-latitude active region, the polar active region, and an inactive region. For most observations, the low-latitude and polar active regions remain stable, as shown by the blue and dashed orange lines in Fig.~\ref{figure2}. However, as discussed in Section~\ref{sec:obs}, a few spectra exhibit significantly stronger emissions (black lines in the upper panels of Fig.~\ref{figure1}), indicating enhanced activity. In these cases, the polar active region requires increased activity, represented by the dotted orange line in Fig.~\ref{figure2}, while the low-latitude active region remains unchanged.
This result of localized polar heating is independently supported by contemporaneous ZDI maps, which show that the polar magnetic field on AD Leo can undergo episodes of strengthening \citep{2023A&A...676A..56B}, providing a physically plausible driver for the enhanced chromospheric heating in that region.

\subsection{Effects of magnetic field strength}\label{sec:magnetic}

\begin{figure*}
  \centering
  \includegraphics[width=0.9\textwidth]{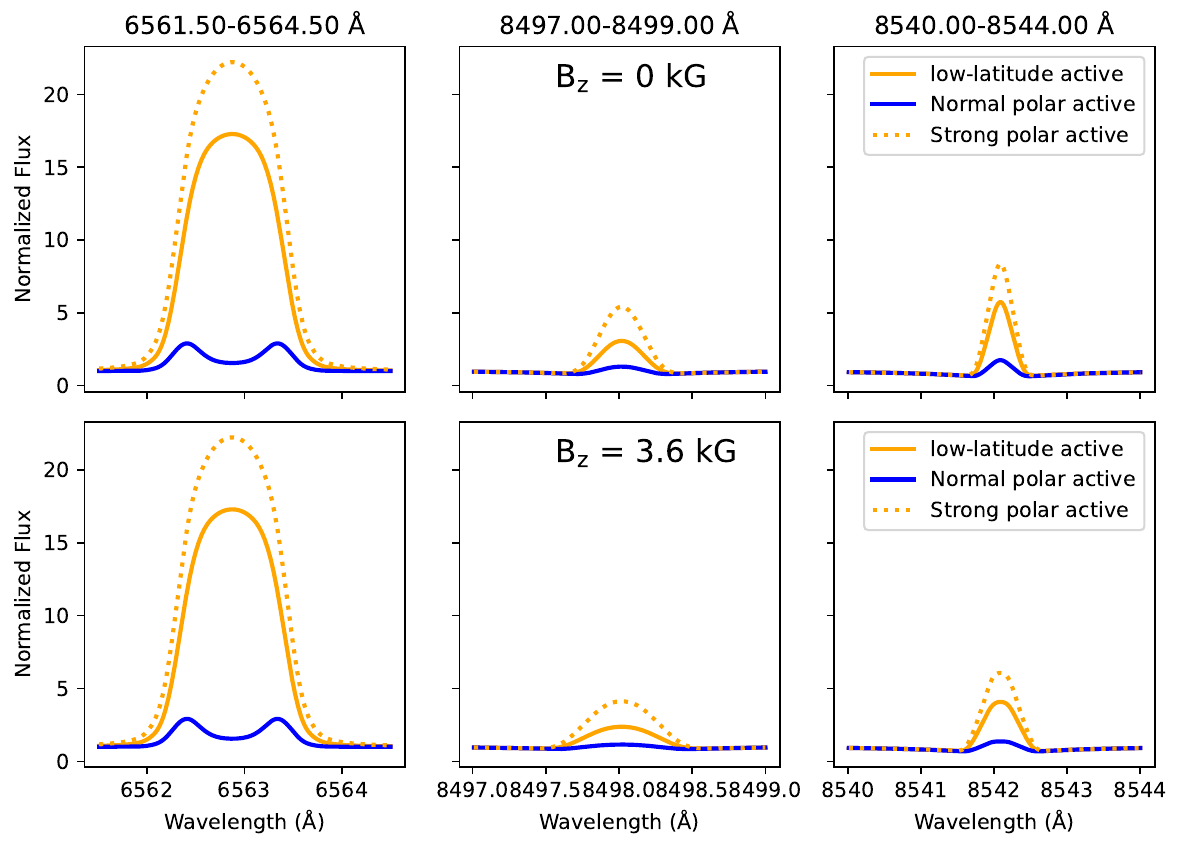}
  \caption{Convolved synthetic profiles for H$\alpha$ (6561.50–6564.50 \AA), \ion{Ca}{II} 8498 \AA (8497.00–8499.00 \AA), and \ion{Ca}{II} 8542 \AA (8540.00–8544.00 \AA) at two magnetic field strengths: $B_z$ = 0.0 kG (left) and 3.6 kG (right). Each panel shows normalized flux versus wavelength for three activity levels: low-latitude active (solid orange), normal polar active (solid blue), and strong polar active (dotted orange).}
  \label{fig:zeeman_effect}
\end{figure*}

Although Zeeman broadening can be modeled by RH1.5D, we deliberately excluded it from our main spectral fitting to maintain focus on the primary goal of constraining the chromospheric thermal structure. Introducing magnetic field strength as an additional free parameter for each atmospheric component would greatly increase model complexity and computational cost, and could introduce significant degeneracies with the temperature and filling factor parameters we aim to determine. To quantitatively assess the potential impact of this simplification, we evaluated the influence of magnetic broadening on our diagnostics by computing synthetic profiles for H\(\alpha\) and \ion{Ca}{II} 8542\,\AA\,with line-of-sight field strengths of 0.0 kG (non-magnetic) and 3.6 kG, matching the unsigned field estimates from Zeeman broadening reported by \citet{2023A&A...676A..56B}. These profiles were generated using the RH1.5D code, in which the magnetic field strength ($B_z$) is included as an input parameter, and were convolved with the CARMENES instrumental profile (R = 94,600), together with rotational broadening for $v\sin i \approx 3$\,km\,s$^{-1}$. After convolution with the CARMENES instrumental profile, the H\(\alpha\) line remains well approximated by a Gaussian function at both tested field strengths, confirming that Zeeman splitting is negligible and justifying our Gaussian fitting procedure. In contrast, the \ion{Ca}{II} IRT lines exhibit significant magnetic broadening, as clearly seen when comparing the 0.0 kG and 3.6 kG cases in Fig. \ref{fig:zeeman_effect}. The 8542 \AA\,transition, in particular, displays a pronounced widening at the higher field strength case. However, even at 3.6\,kG, the impact is limited to this line broadening rather than clearly resolved splitting. These results demonstrate that, within AD Leo's magnetic-field regime, H\(\alpha\) serves as a robust tracer of chromospheric heating, while \ion{Ca}{II} IRT provides a sensitive probe of the surface magnetic flux \citep{1989A&A...225..456S, 1997A&A...326.1135D}.

Figure~\ref{fig:zeeman_effect} compares the convolved synthetic profiles for the non-magnetic (B = 0.0 kG) and strong-field (B = 3.6 kG) cases, using all three model atmospheres: the low-latitude active region, the normal polar active region, and the strong polar active region. Each row corresponds to a different atmospheric model, and the columns display H$\alpha$, \ion{Ca}{II} 8498\,\AA, and \ion{Ca}{II} 8542\,\AA, respectively. The figure confirms that while H$\alpha$ is largely unaffected, the Ca II IRT lines exhibit measurable broadening at 3.6 kG. Across all models and for both field strengths, the profiles remain smooth and symmetric, with no discernible Zeeman splitting after convolution. The broadening manifests as a slight widening without altering the fundamental shape of the line cores.

Therefore, we conclude that while magnetic broadening is measurable for the Ca II IRT lines, it does not significantly impede our ability to derive the chromospheric thermal structure from the line core intensities and overall profiles. The results confirm that the magnetic broadening has a limited influence on the appearance of H$\alpha$ and only introduces subtle broadening, rather than perceptible morphological changes, to the \ion{Ca}{II} IRT lines under the magnetic field strengths typical of AD~Leo. This justifies our model's focus on thermal parameters.

\subsection{Inactive-region contribution}\label{sec:inactive}
\begin{figure*}[htbp]
\includegraphics[width=0.85\hsize]{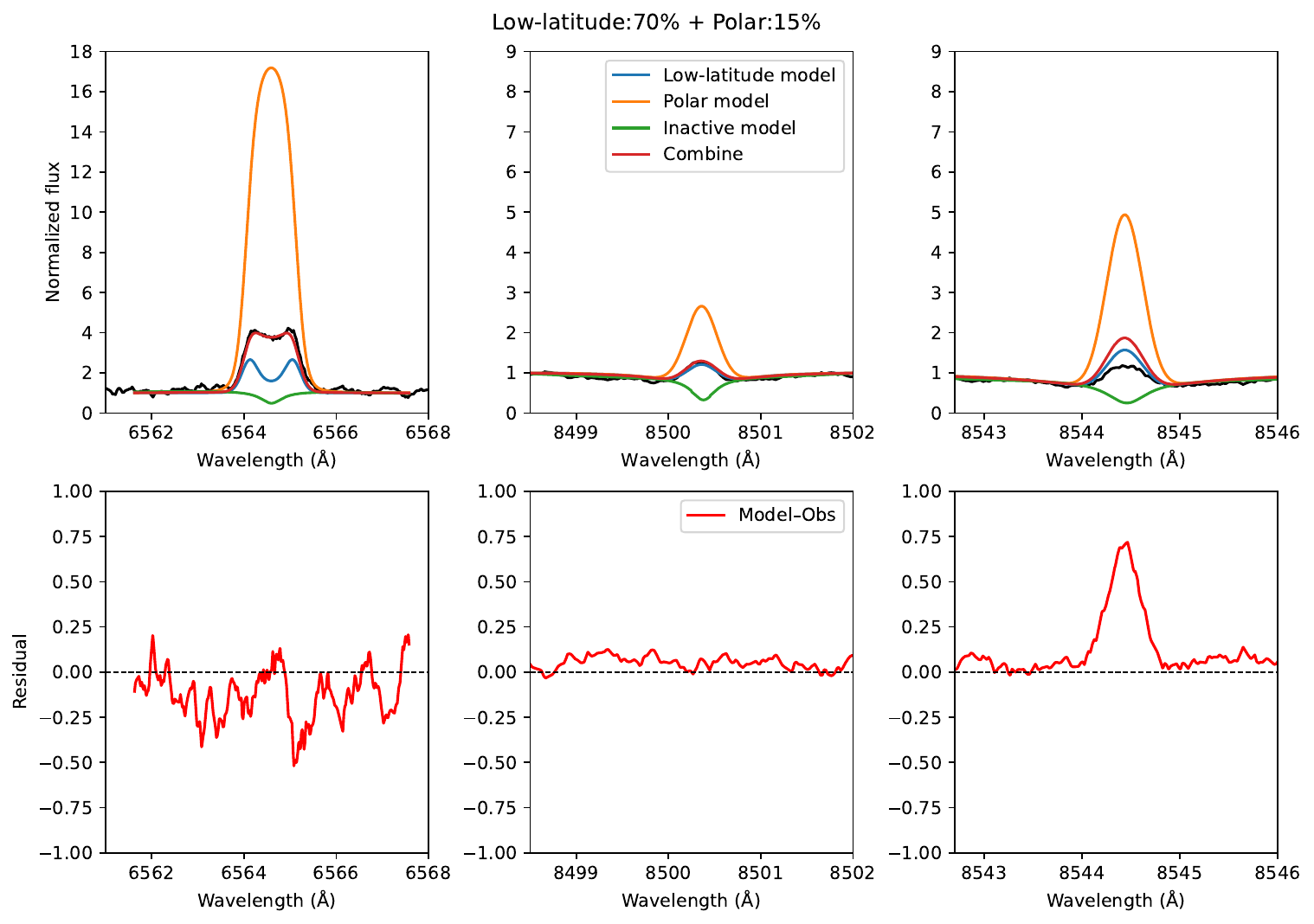}
\includegraphics[width=0.85\hsize]{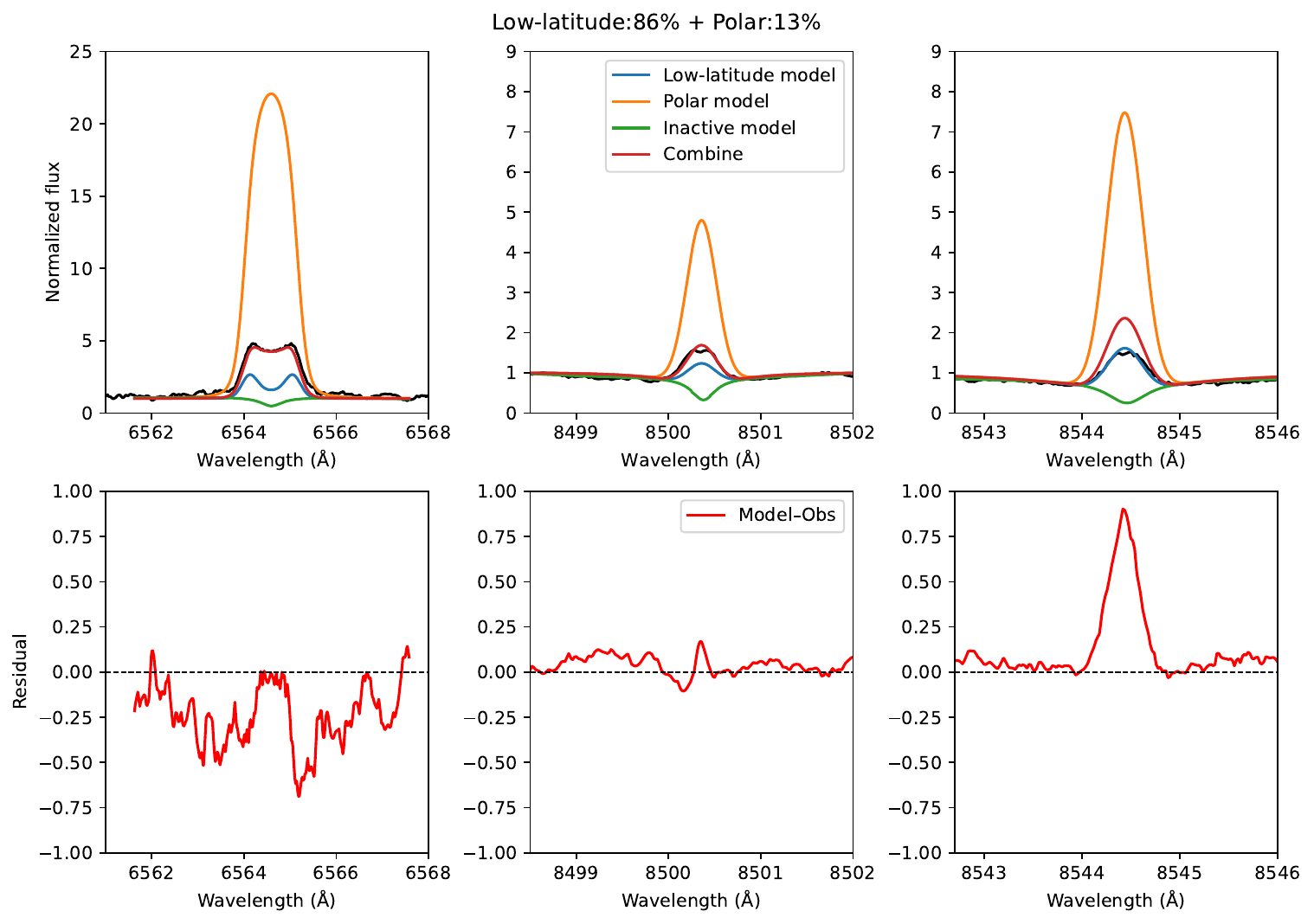}
\caption{Comparison of AD Leo spectra with our multi‐component model: the top two rows correspond to a representative observation fitted with nominal polar and low-latitude active regions plus the inactive component, while the bottom two rows use an enhanced polar active region to match the four high‐flux epochs. Observed data are in black; individual model contributions from the low-latitude and polar active regions are shown in blue and orange, respectively; the quiet‐region template in green; and the composite fit in red, demonstrating accurate reproduction of both line cores and wings. he second and fourth rows in each column show the residuals (model minus observation) in red, highlighting the small deviations between model and data.}

\label{figure3}
\end{figure*}

Although the quiet-region template contributes only 1–15\% of the total flux (Fig.~\ref{figure3}), its inclusion is crucial. Numerically, the low-activity component provides a numerically stable background for the source function that prevents divergence in the MULTI-RH1.5D iterations.
Physically, without this component the synthetic H\(\alpha\) and \ion{Ca}{II} IRT wings become too narrow and the cores overly deep, in conflict with the smooth wings and moderate core depths observed (Fig.~\ref{figure3}, upper panel). High-resolution solar images and Doppler maps of other M dwarfs \citep{2023A&A...676A..56B, 2008MNRAS.390..567M} confirm that chromospheres are inherently patchy, with interspersed active and quiescent areas.

\subsection{Multi-component spectral fits}\label{sec:combination}
Building upon the established chromospheric structures (Section~\ref{sec:structures}) and the contribution from the inactive region (Section~\ref{sec:inactive}), we investigate the spectral line profiles in AD~Leo using our two-component model. This approach was motivated by our previous findings in \citet{2024ApJ...975..133L}, where a single-component model failed to simultaneously fit both H$\alpha$ and \ion{Ca}{II} lines, while a two-component solution provided satisfactory fits. To validate this framework for AD~Leo, we tested three active components via coarse grid search. The results consistently showed negligible filling factors (<1\%) for the third component, while the low-latitude and polar components retained their parameter values. This supports the two-component model as the most parsimonious solution for AD~Leo's chromospheric activity.

We assessed possible degeneracies between filling factors, component temperatures, and the number of active regions by comparing $\chi^2$ values across multiple configurations. The two-component solution consistently yielded the lowest residuals while maintaining physical consistency with the star's large-scale magnetic topology revealed by ZDI results\citep{2023A&A...676A..56B}. Although we cannot exclude the possibility of more complex configurations with additional components, any such components would likely exhibit temperature structures similar to those already identified in our model.

The simplified radiative transfer approach remains justified as instrumental broadening dominates over Zeeman effects in our diagnostic lines for typical field strengths observed in AD~Leo.

Fig.~\ref{figure3} demonstrates the model performance under this magnetic regime. In the top two rows, a typical state is fitted using a low-latitude active region (70\% contribution, blue line), a polar active region (15\%, orange line), and an inactive region (15\%, green line). This combination successfully reproduces the H$\alpha$ line (6563--6568\,\AA) and the bluest \ion{Ca}{II} IRT line at 8500\,\AA, with the combined spectrum (red line) matching the observations (black points). The second and fourth rows in each column display the residuals (model minus observation) in red, highlighting the small deviations between model and data. The fitting was performed over four rotation periods per epoch, with overall phase coverage exceeding one full rotation cycle (see Table~\ref{table:combined}).

In the lower panels, four epochs of enhanced emission (black lines in Fig.~\ref{figure1}) are addressed by using an intensified polar region: the low-latitude region contributes 86\% (blue), the stronger polar region 13\% (orange), and the inactive region only 1\% (green). Although this configuration captures the heightened H$\alpha$ and 8500\,\AA\ emissions, a systematic discrepancy remains in the red \ion{Ca}{II} IRT line at 8544\,\AA, where the model overpredicts the core intensity despite accurate fits elsewhere.

\subsection{Temporal evolution and ZDI correlation}\label{sec:temporal}

Fig.~\ref{figure4} shows the temporal evolution of these filling factors during two observational campaigns. During the 2018 observations (left panel), the low-latitude filling factor ranges between 60\% and 86\%, while the polar component varies between 12\% and 17\%. Although contemporaneous ZDI data are unavailable for this period, we note that \citet{2023A&A...676A..56B} report the ZDI epochs in mid‑2019 (“2019b”) showing an increase in polar radial field strength (see their Fig.~8), which may be connected to the transient enhancements in our polar component during early 2018.

A similar analysis for the 2020 campaign reveals a comparable pattern. For the 2020 observations (right panel), which align with the “2020a” epoch in ZDI maps, our results show consistent behavior: the low-latitude filling factor remains stable at 55–72\%, while the polar component contribution persists at 12–17\%. To place these chromospheric results in the broader context of the star’s magnetic cycle, we compare them with contemporaneous photospheric maps. Fig.~\ref{fig:zdi_maps} (reproduced with permission from \citealt{2023A&A...676A..56B}) shows the corresponding ZDI radial field maps for epochs 2019b and 2020a, illustrating the weakened global dipole and sustained mid-latitude features, consistent with the stable low-latitude component and modest polar-component variations seen in our analysis. In their photospheric analysis, \citet{2023A&A...676A..56B} reported a predominantly poloidal, dipole-dominated large-scale field with indications of long-term evolution possibly linked to a magnetic cycle. While our diagnostics probe the chromosphere, the temporal variability patterns we recover — including the persistence of high-latitude structures and changes in overall activity level — are consistent with the trends they described, suggesting a close connection between large-scale magnetic morphology and chromospheric heating.

The stability of our chromospheric filling factors matches the magnetic topology variations: reduced large‑scale dipolar flux accompanied by persistent low–latitude field concentrations.

Table~\ref{table:combined} summarizes the CARMENES and ZDI epochs side by side, including Julian dates and rotation phase coverage, providing a convenient, epoch-by-epoch reference that facilitates direct comparison of photospheric magnetic maps with our chromospheric diagnostics.

The agreement between spectroscopic filling factors and ZDI magnetic topologies—particularly the correspondence between polar field variations and polar component behavior—strengthens the evidence that large-scale magnetic morphology plays a key role in regulating chromospheric heating \citep{2014ApJ...781L...9B, 2023A&A...676A..56B}.

\subsection{Uncertainty analysis}\label{sec:uncertainty}

The adopted \(1\sigma\) uncertainties on the stellar atmosphere parameters for our low-latitude MARCS-OS model were informed by the typical errors reported in recent K-dwarf analyses. Specifically, \citet{2021A&A...656A.162M} provided the representative uncertainties of \(\sigma_{T_{\rm eff}} = 23\)\,K in effective temperature, \(\sigma_{\log g} = 0.12\)\,dex in surface gravity, and \(\sigma_{\rm [Fe/H]} = 0.12\)\,dex in metallicity. We adopted these values as estimates of plausible parameter uncertainties---not as direct measurements---to assess the sensitivity of our derived filling factors. To propagate these uncertainties, we constructed six perturbed atmosphere models, each offset by \(+1\sigma\) or \(-1\sigma\) in a single parameter while holding the other two fixed at their nominal values. Specifically, we generated models at
\[
\begin{gathered}
(T_{\rm eff} \pm 23\,\mathrm{K},\; \log g,\; {\rm [Fe/H]})\,,\\
(T_{\rm eff},\; \log g \pm 0.12\,\mathrm{dex},\; {\rm [Fe/H]})\,,\\
(T_{\rm eff},\; \log g,\; {\rm [Fe/H]} \pm 0.12\,\mathrm{dex})\,,
\end{gathered}
\]
in addition to the baseline model.

Each perturbed model was then processed through our multi-component spectral fitting routine exactly as for the fiducial case. We re-measured the filling factors \(f_{\rm prim}\) and \(f_{\rm sec}\) for the low-latitude and polar active regions under each perturbation. The maximal absolute deviations of the filling factors, relative to the baseline solution, can be summarized succinctly in prose: perturbations of \(\pm 23\)\,K in \(T_{\rm eff}\) induce shifts up to 2 percent in both \(f_{\rm prim}\) and \(f_{\rm sec}\); variations of \(\pm 0.12\)\,dex in \(\log g\) produce the largest effect, with filling-factor changes up to 15 percent; and metallicity offsets of \(\pm 0.12\)\,dex yield filling-factor shifts not exceeding three percent for either component.

\subsection{Implications for M dwarf activity}\label{sec:implications}

Our two-component chromospheric model—featuring a stable, VAL-C-like polar structure and a dynamic equatorial region—reveals critical insights into M dwarf heating mechanisms. The persistent polar component (contributing 12–17\% of flux) indicates large-scale, dipole-anchored heating maintained over multi-year timescales, driven by global dynamo processes \citep{2012PASP..124..545H}. In contrast, the variable equatorial component (55–86\% flux) reflects localized, flare-associated heating modulated by short-term magnetic reconnection events \citep{2023A&A...674A.110C}.

\begin{figure}
  \centering
  \includegraphics[width=1\columnwidth]{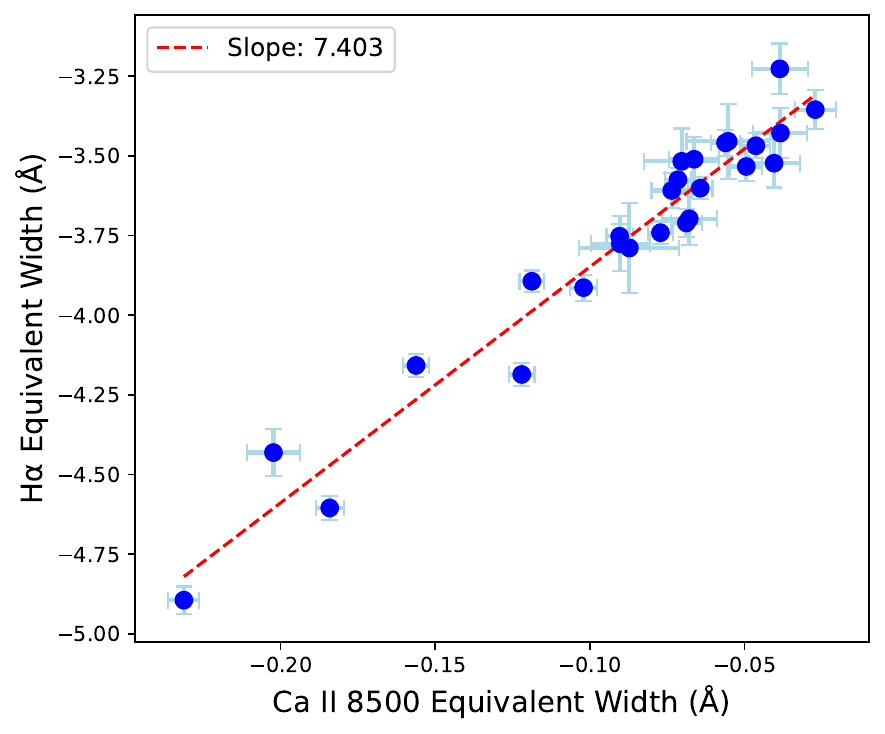}
  \caption{Scatter plot of H$\alpha$ versus \ion{Ca}{II} IRT equivalent widths for AD Leo.}
  \label{fig:ha_ca_corr}
\end{figure}

This structural dichotomy carries significant implications. First, regarding stellar activity cycles, in our CARMENES time-series covering rotational and seasonal timescales, H$\alpha$ and \ion{Ca}{II} 8500\,Å equivalent widths exhibit a positive trend (slope = 7.403; Fig.~\ref{fig:ha_ca_corr}), which reflects co-variation on short timescales ($\sim$1-month) and differs fundamentally in timescale from the $\sim$7-year magnetic cycle examined by \citet{2020A&A...642A..53D}. They reported anti-correlation emerges only on this long timescale, whereas our result pertains to short-term variability, so the two findings are not necessarily contradictory.

We have tested the sufficiency of the two-component model by introducing a third atmospheric component: in every test, the filling factor of the third-active component component converged to below 1\%, and no improvement in the minimum $\chi^2$ was achieved compared to the two-component fit. This confirms the statistical and physical parsimony of our chosen model, although we acknowledge that absolute latitudinal placement remains model-dependent within our 1.5D framework.

Second, for exoplanet space weather, the stable polar component likely generates persistent XUV/EUV backgrounds while episodic equatorial flares drive transient high-energy particle fluxes \citep{2020A&A...637A..13M}. Third, in chromospheric modeling, the consistent success of this framework for both moderately active (G\,80-21) and highly active stars (AD Leo) suggests broad applicability across the M dwarf activity sequence.

Future investigations should prioritize three directions: coordinated \text{time-resolved} spectroscopy (H$\alpha$, \ion{Ca}{II} IRT) with ZDI to resolve magnetic–heating coupling, i.e., the spatial association between stronger magnetic fields in the ZDI maps and regions of elevated chromospheric temperature required by our modeling; extension to ultracool dwarfs where convective suppression alters dynamo behavior; and quantitative modeling of energetic particle fluxes from equatorial activity belts.

\section{Conclusions}\label{sec:con}
In this study, we explored the chromospheric structure of the active M dwarf AD Leo by combining high-resolution spectroscopic observations with contemporaneous magnetic field maps derived from ZDI. The two-component model we employed—featuring a dynamic low-latitude region and a relatively stable polar component that can be episodically enhanced—was able to reproduce key features in the H$\alpha$ and \ion{Ca}{II} infrared triplet lines across multiple epochs. Crucially, the ZDI maps served as an independent, qualitative reference for the large-scale magnetic topology and are not used as input constraints in the model optimization.

Our analysis suggests that the low-latitude region shows notable variability in surface coverage (55--86\%), while maintaining a stable thermal structure. Conversely, the polar region appears more stable in area (12--17\%) but requires modifications in temperature during high-activity phases. 

By incorporating both spectroscopic and qualitative magnetic context, this approach provides a physically motivated framework to investigate spatial and temporal inhomogeneities in M dwarf chromospheres. In particular, we find that the evolution of the filling factors across different epochs shows qualitative consistency with the large-scale magnetic topology changes seen in the ZDI maps, such as the weakening of the dipolar field and the increasing obliquity reported by \citet{2023A&A...676A..56B}. This consistency, discussed in detail in Section~\ref{sec:result} and shown in Fig.~\ref{fig:zdi_maps}, provides independent support for the inferred spatial configuration of the chromospheric components.

Although limited to a single target, the results illustrate the potential of combining ZDI as an independent diagnostic with non-LTE spectral modeling to study magnetic activity in low-mass stars.

Future work could extend this methodology to a larger sample of active M dwarfs, enabling a more systematic investigation of how magnetic topology influences chromospheric behavior and, by extension, the space plasma environment surrounding habitable-zone planets.

\begin{acknowledgements}
We thank the referee for his/her valuable suggestions and insightful comments, which greatly helped to improve the quality of this manuscript.
This research is supported by the International Partnership Program of the Chinese Academy of Sciences (Grant No. 178GJHZ2022047GC). Additional support was provided by the National Natural Science Foundation of China under Grant Nos. 12588202, 12090040/4, 12373036, 12022304, 11973052, and 12173058; the National Key R\&D Program of China (Grant No. 2019YFA0405502); and the Scientific Instrument Developing Project of the Chinese Academy of Sciences (Grant No. ZDKYYQ20220009).
We also thank M. S. Giampapa for his insightful comments and suggestions during the preparation of an earlier version of this work.
\end{acknowledgements}

\bibliography{ref}{}
\bibliographystyle{aa}

\twocolumn 
\appendix 
\section{Model grid construction}

\begin{table}[h!]
\setlength{\tabcolsep}{3pt}
\centering
\caption{Parameters Used in Model Grid Generation (Part 1)}
\label{table:set}
\begin{tabular}{ccccccc}
\hline\hline
$T_{\min}$ & $m_{\min}$ & $T_{\rm mid}$ & $m_{\rm mid}$ & $T_{\rm top}$ & $m_{\rm top}$ & $\rm{grad}_{\rm TR}$ \\
(K) & (dex) & (K) & (dex) & (K) & (dex) & \\
\hline
2558 & $-$2.0 & 6000 & $-$2.5 & 7500 & $-$4.0 & 31622780 \\
2591 & $-$1.5 & 5000 & $-$2.5 & 7500 & $-$3.0 & 31622780 \\
2558 & $-$2.0 & 5000 & $-$2.5 & 8000 & $-$3.0 & 31622780 \\
2591 & $-$1.5 & 6000 & $-$3.0 & 6500 & $-$3.5 & 31622780 \\
2531 & $-$2.5 & 6000 & $-$3.0 & 7500 & $-$3.5 & 31622780 \\
2558 & $-$2.0 & 6000 & $-$2.5 & 7500 & $-$3.0 & 31622780 \\
2558 & $-$2.0 & 6000 & $-$3.0 & 7500 & $-$3.5 & 31622780 \\
2591 & $-$1.5 & 7000 & $-$2.0 & 8000 & $-$2.5 & 31622780 \\
2591 & $-$1.5 & 6000 & $-$2.5 & 8000 & $-$3.0 & 31622780 \\
2591 & $-$1.5 & 5000 & $-$2.5 & 7500 & $-$4.0 & 31622780 \\
2591 & $-$1.5 & 5000 & $-$2.0 & 6500 & $-$4.0 & 31622780 \\
2591 & $-$1.5 & 6000 & $-$2.0 & 8000 & $-$3.5 & 31622780 \\
2591 & $-$1.5 & 7000 & $-$2.0 & 7500 & $-$3.0 & 31622780 \\
2558 & $-$2.0 & 5000 & $-$2.5 & 7000 & $-$3.5 & 31622780 \\
2591 & $-$1.5 & 6000 & $-$2.5 & 8000 & $-$3.5 & 31622780 \\
2531 & $-$2.5 & 6000 & $-$3.0 & 8000 & $-$3.5 & 31622780 \\
2591 & $-$1.5 & 7000 & $-$3.0 & 7500 & $-$3.5 & 31622780 \\
2591 & $-$1.5 & 6000 & $-$2.0 & 7000 & $-$4.0 & 31622780 \\
2591 & $-$1.5 & 6000 & $-$2.5 & 6500 & $-$3.5 & 31622780 \\
2558 & $-$2.0 & 6000 & $-$2.5 & 8000 & $-$4.0 & 31622780 \\
2591 & $-$1.5 & 5000 & $-$2.5 & 6500 & $-$3.0 & 31622780 \\
2591 & $-$1.5 & 5000 & $-$2.0 & 7500 & $-$3.5 & 31622780 \\
2591 & $-$1.5 & 6000 & $-$2.0 & 8000 & $-$3.0 & 31622780 \\
2558 & $-$2.0 & 7000 & $-$2.5 & 7500 & $-$3.0 & 31622780 \\
2591 & $-$1.5 & 7000 & $-$2.5 & 8000 & $-$3.5 & 31622780 \\
2591 & $-$1.5 & 6000 & $-$2.0 & 7500 & $-$2.5 & 31622780 \\
2558 & $-$2.0 & 6000 & $-$3.0 & 8000 & $-$3.5 & 31622780 \\
2591 & $-$1.5 & 5000 & $-$2.0 & 7000 & $-$3.0 & 31622780 \\
2558 & $-$2.0 & 5000 & $-$2.5 & 6500 & $-$3.0 & 31622780 \\
2558 & $-$2.0 & 7000 & $-$2.5 & 7500 & $-$3.5 & 31622780 \\
2591 & $-$1.5 & 6000 & $-$2.0 & 7000 & $-$3.0 & 31622780 \\
2591 & $-$1.5 & 5000 & $-$2.5 & 7500 & $-$3.5 & 31622780 \\
2591 & $-$1.5 & 5000 & $-$2.0 & 6500 & $-$3.0 & 31622780 \\
2591 & $-$1.5 & 5000 & $-$2.5 & 7000 & $-$3.0 & 31622780 \\
2558 & $-$2.0 & 7000 & $-$2.5 & 8000 & $-$3.0 & 31622780 \\
2558 & $-$2.0 & 7000 & $-$3.0 & 8000 & $-$4.0 & 31622780 \\
2591 & $-$1.5 & 7000 & $-$2.0 & 7500 & $-$4.0 & 31622780 \\
2591 & $-$1.5 & 7000 & $-$3.0 & 8000 & $-$3.5 & 31622780 \\
2531 & $-$2.5 & 7000 & $-$3.0 & 8000 & $-$4.0 & 31622780 \\
2558 & $-$2.0 & 5000 & $-$2.5 & 8000 & $-$3.5 & 31622780 \\
2558 & $-$2.0 & 6000 & $-$2.5 & 7000 & $-$3.5 & 31622780 \\
2591 & $-$1.5 & 6000 & $-$2.5 & 7500 & $-$3.5 & 31622780 \\
2591 & $-$1.5 & 5000 & $-$2.0 & 8000 & $-$2.5 & 31622780 \\
2591 & $-$1.5 & 7000 & $-$3.0 & 7500 & $-$4.0 & 31622780 \\
2591 & $-$1.5 & 6000 & $-$2.0 & 7000 & $-$3.5 & 31622780 \\
2591 & $-$1.5 & 5000 & $-$2.0 & 7500 & $-$4.0 & 31622780 \\
2591 & $-$1.5 & 6000 & $-$2.5 & 7000 & $-$3.0 & 31622780 \\
2558 & $-$2.0 & 6000 & $-$2.5 & 6500 & $-$3.0 & 31622780 \\
2591 & $-$1.5 & 7000 & $-$2.5 & 7500 & $-$4.0 & 31622780 \\
2591 & $-$1.5 & 7000 & $-$2.0 & 7500 & $-$2.5 & 31622780 \\
2591 & $-$1.5 & 6000 & $-$2.0 & 6500 & $-$3.0 & 31622780 \\
2558 & $-$2.0 & 5000 & $-$2.5 & 8000 & $-$4.0 & 31622780 \\
2591 & $-$1.5 & 5000 & $-$2.0 & 7500 & $-$2.5 & 31622780 \\
2591 & $-$1.5 & 5000 & $-$2.0 & 8000 & $-$3.0 & 31622780 \\
2591 & $-$1.5 & 6000 & $-$3.0 & 7500 & $-$4.0 & 31622780 \\
2558 & $-$2.0 & 6000 & $-$2.5 & 7000 & $-$4.0 & 31622780 \\
\hline
\end{tabular}
\end{table}

\begin{table}[h!]
\setlength{\tabcolsep}{3pt}
\centering
\caption{Parameters Used in Model Grid Generation (Part 2)}
\begin{tabular}{ccccccc}
\hline\hline
$T_{\min}$ & $m_{\min}$ & $T_{\rm mid}$ & $m_{\rm mid}$ & $T_{\rm top}$ & $m_{\rm top}$ & $\rm{grad}_{\rm TR}$ \\
(K) & (dex) & (K) & (dex) & (K) & (dex) & \\
\hline

2591 & $-$1.5 & 6000 & $-$2.5 & 6500 & $-$4.0 & 31622780 \\
2591 & $-$1.5 & 5000 & $-$2.0 & 8000 & $-$4.0 & 31622780 \\
2591 & $-$1.5 & 6000 & $-$2.0 & 6500 & $-$3.5 & 31622780 \\
2558 & $-$2.0 & 6000 & $-$2.5 & 7000 & $-$3.0 & 31622780 \\
2591 & $-$1.5 & 6000 & $-$2.5 & 6500 & $-$3.0 & 31622780 \\
2591 & $-$1.5 & 7000 & $-$2.0 & 8000 & $-$3.0 & 31622780 \\
2591 & $-$1.5 & 6000 & $-$3.0 & 7500 & $-$3.5 & 31622780 \\
2591 & $-$1.5 & 6000 & $-$3.0 & 8000 & $-$4.0 & 31622780 \\
2591 & $-$1.5 & 5000 & $-$2.0 & 7500 & $-$3.0 & 31622780 \\
2591 & $-$1.5 & 6000 & $-$2.5 & 7000 & $-$4.0 & 31622780 \\
2558 & $-$2.0 & 6000 & $-$2.5 & 6500 & $-$4.0 & 31622780 \\
2558 & $-$2.0 & 7000 & $-$2.5 & 8000 & $-$4.0 & 31622780 \\
2591 & $-$1.5 & 7000 & $-$2.5 & 7500 & $-$3.5 & 31622780 \\
2591 & $-$1.5 & 5000 & $-$2.0 & 8000 & $-$3.5 & 31622780 \\
2591 & $-$1.5 & 5000 & $-$2.5 & 7000 & $-$3.5 & 31622780 \\
2591 & $-$1.5 & 6000 & $-$2.0 & 7500 & $-$3.0 & 31622780 \\
2591 & $-$1.5 & 7000 & $-$2.5 & 8000 & $-$3.0 & 31622780 \\
2591 & $-$1.5 & 7000 & $-$2.0 & 8000 & $-$4.0 & 31622780 \\
2558 & $-$2.0 & 7000 & $-$3.0 & 7500 & $-$3.5 & 31622780 \\
2591 & $-$1.5 & 6000 & $-$2.0 & 8000 & $-$4.0 & 31622780 \\
2531 & $-$2.5 & 7000 & $-$3.0 & 7500 & $-$3.5 & 31622780 \\
2558 & $-$2.0 & 5000 & $-$2.5 & 6500 & $-$3.0 & 31622780 \\
2591 & $-$1.5 & 6000 & $-$2.5 & 7000 & $-$3.5 & 31622780 \\
2558 & $-$2.0 & 6000 & $-$2.5 & 7500 & $-$3.5 & 31622780 \\
2591 & $-$1.5 & 6000 & $-$3.0 & 8000 & $-$3.5 & 31622780 \\
2591 & $-$1.5 & 6000 & $-$2.0 & 6500 & $-$4.0 & 31622780 \\
2558 & $-$2.0 & 5000 & $-$2.5 & 7500 & $-$3.5 & 31622780 \\
2558 & $-$2.0 & 6000 & $-$2.5 & 8000 & $-$3.5 & 31622780 \\
2531 & $-$2.5 & 7000 & $-$3.0 & 7500 & $-$4.0 & 31622780 \\
2591 & $-$1.5 & 6000 & $-$2.0 & 7500 & $-$4.0 & 31622780 \\
2591 & $-$1.5 & 7000 & $-$3.0 & 8000 & $-$4.0 & 31622780 \\
2531 & $-$2.5 & 7000 & $-$3.0 & 8000 & $-$3.5 & 31622780 \\
2558 & $-$2.0 & 6000 & $-$2.5 & 6500 & $-$3.5 & 31622780 \\
2558 & $-$2.0 & 5000 & $-$2.5 & 7000 & $-$3.0 & 31622780 \\
2591 & $-$1.5 & 6000 & $-$2.5 & 8000 & $-$4.0 & 31622780 \\
2591 & $-$1.5 & 5000 & $-$2.0 & 7000 & $-$3.5 & 31622780 \\
2591 & $-$1.5 & 5000 & $-$2.5 & 8000 & $-$3.5 & 31622780 \\
2591 & $-$1.5 & 7000 & $-$2.0 & 7500 & $-$3.5 & 31622780 \\
2591 & $-$1.5 & 7000 & $-$2.5 & 7500 & $-$3.0 & 31622780 \\
2558 & $-$2.0 & 7000 & $-$2.5 & 8000 & $-$3.5 & 31622780 \\
2558 & $-$2.0 & 7000 & $-$3.0 & 8000 & $-$3.5 & 31622780 \\
2558 & $-$2.0 & 7000 & $-$3.0 & 7500 & $-$4.0 & 31622780 \\
2591 & $-$1.5 & 7000 & $-$3.0 & 8000 & $-$4.0 & 31622780 \\
2591 & $-$1.5 & 6000 & $-$3.0 & 7000 & $-$4.0 & 31622780 \\
2591 & $-$1.5 & 6000 & $-$2.5 & 7500 & $-$4.0 & 31622780 \\
2558 & $-$2.0 & 5000 & $-$2.5 & 7500 & $-$3.0 & 31622780 \\
2591 & $-$1.5 & 5000 & $-$2.5 & 8000 & $-$3.0 & 31622780 \\
2591 & $-$1.5 & 6000 & $-$3.0 & 7000 & $-$3.5 & 31622780 \\
2591 & $-$1.5 & 6000 & $-$2.5 & 7500 & $-$3.0 & 31622780 \\
2558 & $-$2.0 & 6000 & $-$2.5 & 8000 & $-$3.0 & 31622780 \\
2591 & $-$1.5 & 6000 & $-$2.0 & 7500 & $-$3.5 & 31622780 \\
2591 & $-$1.5 & 5000 & $-$2.0 & 7000 & $-$4.0 & 31622780 \\
2591 & $-$1.5 & 5000 & $-$2.5 & 8000 & $-$4.0 & 31622780 \\
2591 & $-$1.5 & 5000 & $-$2.0 & 6500 & $-$3.5 & 31622780 \\
2591 & $-$1.5 & 7000 & $-$2.0 & 8000 & $-$3.5 & 31622780 \\
\hline
\end{tabular}
\end{table}

\end{document}